\documentclass[%
 reprint,
longbibliography,
 amsmath,amssymb,
 aps,
prb,
]{revtex4-1}
\setcitestyle{numbers,square}

\usepackage{braket}
\usepackage{float}
\usepackage{placeins}
\usepackage{graphicx}
\usepackage{dcolumn}
\usepackage{bm}
\usepackage[normalem]{ulem} 
\usepackage{xcolor}
\usepackage{hyperref}
\hypersetup{
    colorlinks,%
    citecolor=blue,%
    linkcolor=blue,%
    urlcolor=blue
}

\newcommand{\orcid}[1]{\href{https://orcid.org/#1}{\includegraphics[width=8pt]{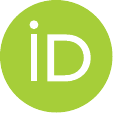}}}

\begin{document}

\title{Reentrant topological phase in half-Heusler compounds}

\author{A. L. Araújo\orcid{0000-0002-6835-6113}}
\email{augusto.araujo@ilum.cnpem.br}
\affiliation{Ilum School of Science, Brazilian Center for Research in Energy and Materials (CNPEM), Campinas, SP, Brazil}

\author{F. Crasto de Lima\orcid{0000-0002-2937-2620}} 
\email{felipe.lima@ilum.cnpem.br}
\affiliation{Ilum School of Science, Brazilian Center for Research in Energy and Materials (CNPEM), Campinas, SP, Brazil}

\author{A. Fazzio\orcid{0000-0001-5384-7676}}
\email{adalberto.fazzio@ilum.cnpem.br}
\affiliation{Ilum School of Science, Brazilian Center for Research in Energy and Materials (CNPEM), Campinas, SP, Brazil}

\date{\today}

\begin{abstract}

Half-Heusler compounds are known for their various compositions and multifunctional properties including topological phases. In this study, we investigate the topological classification of this class of materials based on the ordering of the $\Gamma_6$, $\Gamma_8$ states, and a previously overlooked $\Gamma_6^*$ state, during an adiabatic expansion process. Using first-principles calculations based on density functional theory, we observed that the non-trivial topology is governed by a three-band mechanism. We provide a simple model derived from $k\cdot p$ Hamiltonian that interprets the topological phase in half-Heusler systems. Additionally, we explore the robustness of the topological phase under tension and a new perspective on the topological nature of half-Heusler compounds.

\end{abstract}

\maketitle

\section{Introduction}
\label{sec:introduction}

\begin{figure}[t]
    \includegraphics[width=\columnwidth]{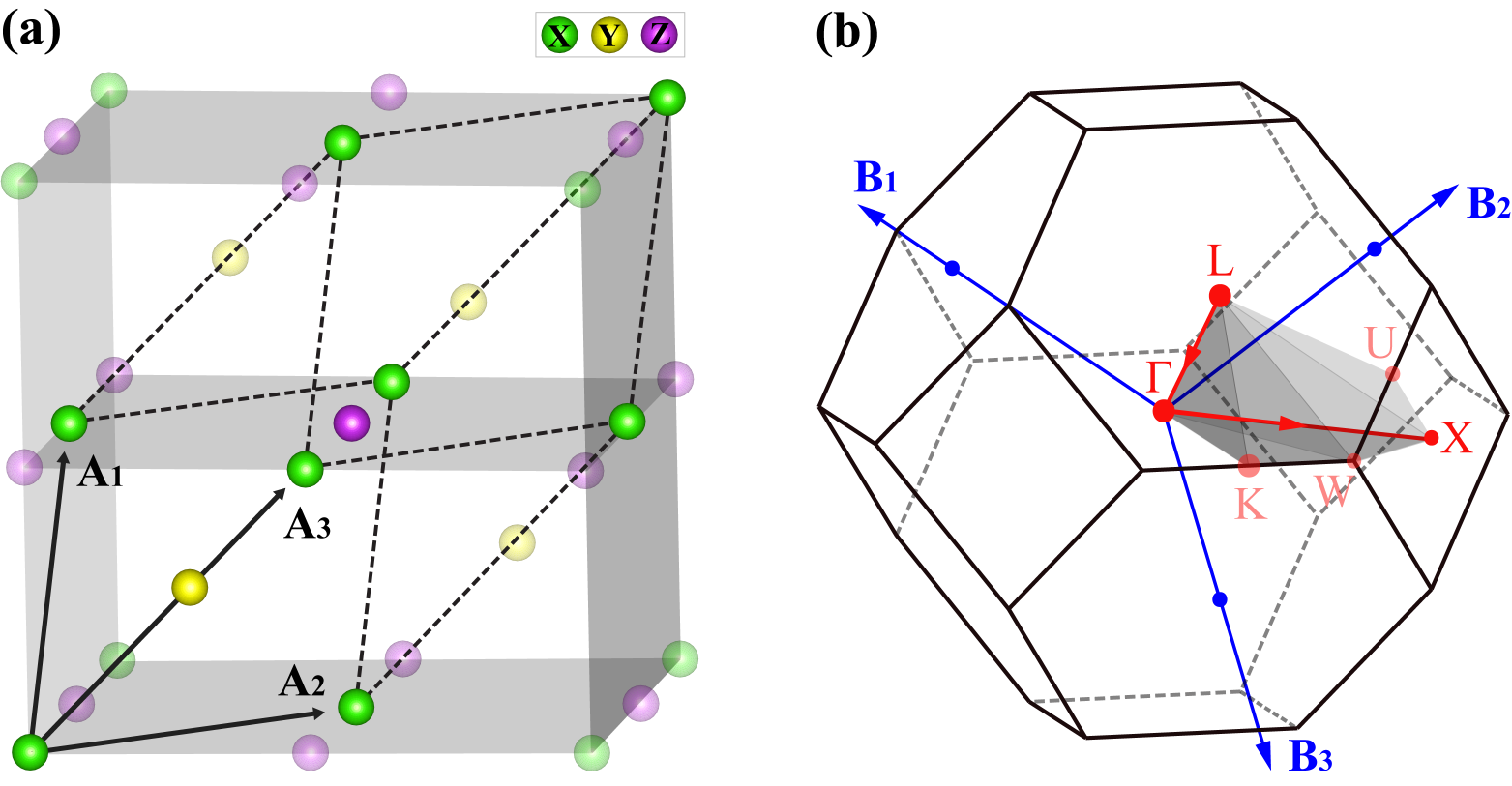}
    \caption{\label{fig:Lattice} (a) Crystal lattice of half-Heusler compounds and (b) corresponding 1st BZ where the points of high symmetry are highlighted as well as the $k$-path taken to plot the band structure.}
\end{figure}

In the quest for topological insulator systems, a key point is the capability to tune the material's properties. This tuning is essential to allow the design of future devices. Materials, named half-Heusler (hH), are ternary with the stoichiometric formulation XYZ [Fig.~\ref{fig:Lattice}(a)], encompassing a wide variety of compositions and functionalities \citep{kawasaki2022full, lin2010, feng2010, xiao2010}. Structurally, these materials crystallize in the F-43m space group and are often compared to the binary zinc-blend III-V semiconductors. One of the main characteristics of Heusler compounds is the multifunctional and tunable nature of their electronic structure. For the same crystal structure, different combinations of atomic species provide a range of emergent properties such as semimetallicity \citep{xia2001}, thermoelectricity \citep{lin2010, felser2007, mastronardi1999, jung2001}, ferromagnetism \citep{katsnelson2008}, superconductivity \citep{goll2008, winterlik2009}, and semiconductivity \citep{kandpal2006}; enabling the creation of materials with combined properties such as topological superconductors \citep{butch2011}. The interaction between the multifunctional electronic nature and topological ordering makes half-Heusler compounds an ideal platform for realizing new topological phases and effects across a wide range of applications.

Numerous theoretical \citep{lin2010, feng2010half, chadov2010t, xiao2010} and experimental \citep{logan2016, chen2017} studies have reported the presence of topological phases in various half-Heusler compounds, where the combination of strong spin-orbit coupling (SOC) and crystal field splitting leads to the inversion of $\Gamma_6$ and $\Gamma_8$ bands around the Fermi level. This explanation is based on the original theoretical prediction of 2D topological insulators, characterized by the $\Gamma_6$/$\Gamma_8$ ordering in the HgTe/CdTe quantum well \cite{bernevig2006,fu2007}. However, it still lacks spectroscopic insight into the corresponding band inversion in this family \cite{PRBsouza2023}. Here fully understanding the nature of the band inversion, can lead both to the arising properties \cite{PhysRevMaterials.7.104203} and its manipulation \cite{Wollmann2017, PhysRevMaterials.7.024802, Shirokura2022, PhysRevMaterials.5.124207}. Particularly, strain engineering in half-Heusler materials is being explored in order to tune topological properties \cite{Bhardwaj2021} with an experimental approach indicating strains over $10\%$ \cite{Du2023}.

In this study, we investigated the topological classification of half-Heusler compounds XYZ hH [X=Lu, Y, Sc; Y=Pt, Au, Ni; and Z=Bi, Sb, Sn], based on the relative ordering of the $\Gamma_6$, $\Gamma_8$, and a previous overlooked $\Gamma_6^*$ states, when strain/pressure is applied to the system. With density functional theory (DFT) calculations we have computed the Z$_2$ topological index, and discuss the orbital origin of the topological phase. We observe that the non-trivial topology is governed by a double band inversion mechanism, providing robustness to the non-trivial phase of this class of materials. This study provides a new perspective on the topological nature of half-Heusler compounds.

\section{METHODS}
\label{sec:methods}

\subsection{Density Functional Theory}

First-principles calculations were performed using DFT \cite{dft1, dft2} within the generalized gradient approximation for the exchange-correlation functional, employing the Perdew-Burke-Ernzerhof parameterization \cite{PBE}. A fully relativistic pseudopotential within the projector augmented-wave (PAW) method \cite{PRBblochl1994, Corso2010} was used in the non-collinear spin-DFT formalism self-consistently. We used the VASP (Vienna ab initio Simulation Package) \cite{vasp1, vasp2} and QE (Quantum Espresso) \cite{Giannozzi2009, Giannozzi2017} DFT packages with a plane-wave basis set defined with a cut-off energy of 400 eV. The Brillouin zone was sampled with a 25x25x25 $k$-point mesh such that the total energy converged within the meV scale. To accurately describe the strong relativistic effect, we included spin-orbit coupling. The analysis of the DFT results, including band structure plots and projections of the atomic orbital contribution, was performed using the VASProcar post-processing code \citep{vasprocar}.

\subsection{Topological Classification}

For the topological classification of half-Heusler compounds, we employed two distinct approaches. The primary approach was topological quantum chemistry \citep{TQC1} via the VASP2Trace code \citep{TQC2}, which establishes a correlation between the material's topological ordering and the elementary band representations (EBRs). This approach addresses the building blocks of the bands through symmetry analysis and the distribution of atomic orbitals, as well as the compatibility relations of how the valence states connect across the Brillouin zone. Additionally, we calculated the Z$_2$ topological index using the WannierTools package \citep{WannierTools}, which analyzes the electronic structure by implementing the recursive Green's function method in a Tight-Binding (TB) Hamiltonian matrix obtained through the Wannier90 package \citep{Wannier90}. To obtain the TB Hamiltonians, we mapped the Bloch functions from the DFT calculation via QE onto a reduced basis set, consisting of 64 maximally localized Wannier functions, using the predominant atomic orbitals of the valence states as a basis.

\subsection{Hamiltonian k$\cdot$p}

To obtain the $k \cdot p$ Hamiltonian, we utilized the method of invariants \citep{Winkler2003} implemented in the Python packages DFT2kp \citep{dft2kp} and QSYMM \citep{qsymm}, to derive the most general form of $H \equiv H(k)$ as a power expansion of $k$, which is invariant under all the symmetries of the $T_d$ point group. For this purpose, we defined the matrix representation of the generators of the $T_d$ group, as well as the time-reversal symmetry $\tau$, in terms of the basis functions related to the states of interest $\Gamma_6$, $\Gamma_6^{*}$, and $\Gamma_8$, leading to an expansion of the well-known Kane model with the inclusion of the high-energy $\Gamma_6^{*}$ state. In order to compact the notation, we write the matrices below in terms of the Pauli matrices $\sigma_{x,y,z}$ and the 2x2 identity matrix $\sigma_0$, in addition to the respective Kronecker products $\Lambda_{i,j} = \sigma_i \otimes \sigma_j$.

To find the effective Hamiltonian, we start with the most general expression up to $n$-order in $k_x$, $k_y$, and $k_z$,
\begin{equation}
H_{\Gamma}^{n} = \sum_{i=0}^{n}\sum_{j=0}^{n}\sum_{m=0}^{n}{Q_{i,j.m}k_x^ik_y^jk_z^m},
\end{equation}
where each $Q_{i,j.m}$ is a general Hermitian $8\times 8 $ matrix, and $k$ is deviating from the $\Gamma$ point of the BZ. Requiring that $H_{\Gamma}$ commutes with all the symmetry elements of $T_d$ and $\tau$, we obtain the following effective Hamiltonian up to first order in $k$. Using the orbital ordering, for s orbitals of the X and Y atoms and p orbitals for the Z atoms $\{ |{\rm Xs}\rangle,\,|{\rm Ys} \rangle,\, |{\rm Zp_x} \rangle,\, |{\rm Zp_y} \rangle,\,|{\rm Zp_z} \rangle \} $, with $|{\rm Xs}\rangle = |{\rm Xs},\uparrow \rangle \otimes |{\rm Xs}, \downarrow \rangle$, we can write
\begin{equation}
H_{k\cdot p} = \begin{pmatrix}
               M_{6} & V_{6{6^{*}}} & V_{86} \\
               V^{\dagger}_{6{6^{*}}} & M_{6^*} & V_{86^*} \\
               V^{\dagger}_{86} & V^{\dagger}_{86^*} & M_{8}
              \end{pmatrix}  \label{eq:kp-ham}
\end{equation}
With each term being related to the $\sigma_i$ and $\Lambda_{i,j}$, where $c_n$ are constant coefficients:
\begin{gather*}
M_{6} = c_{0}\sigma_{0}; \hspace{0.5cm} M_{{6^{*}}} = c_{2}\sigma_{0}; \hspace{0.5cm}  V_{6{6^{*}}} = c_{1}\sigma_{0}
\\ \\
M_{8} = c_{3}\Lambda_{0,0} - c_{6}(\Lambda_{z,x}k_x + \Lambda_{z,y}k_y)  \hspace{0.25cm} + \\ c_{6}{\frac{1}{\sqrt{3}}}(-\Lambda_{x,x}k_x + \Lambda_{x,y}k_y + 2\Lambda_{x,z}k_z)
\\ \\
V_{86} = c_{4}V_{86}; \hspace{0.5cm} V_{8{6^*}} = c_{5}V_{86}
\\
V_{86} = \begin{pmatrix}
          [-i\sigma_{z}k_x + \sigma_{0}k_y] & {\frac{1}{\sqrt{3}}}[i\sigma_{z}k_x + \sigma_{0}k_y + 2i\sigma_{x}k_z] \\
         \end{pmatrix}  
\end{gather*}

\begin{figure*}[t]
\includegraphics[width=2.07\columnwidth]{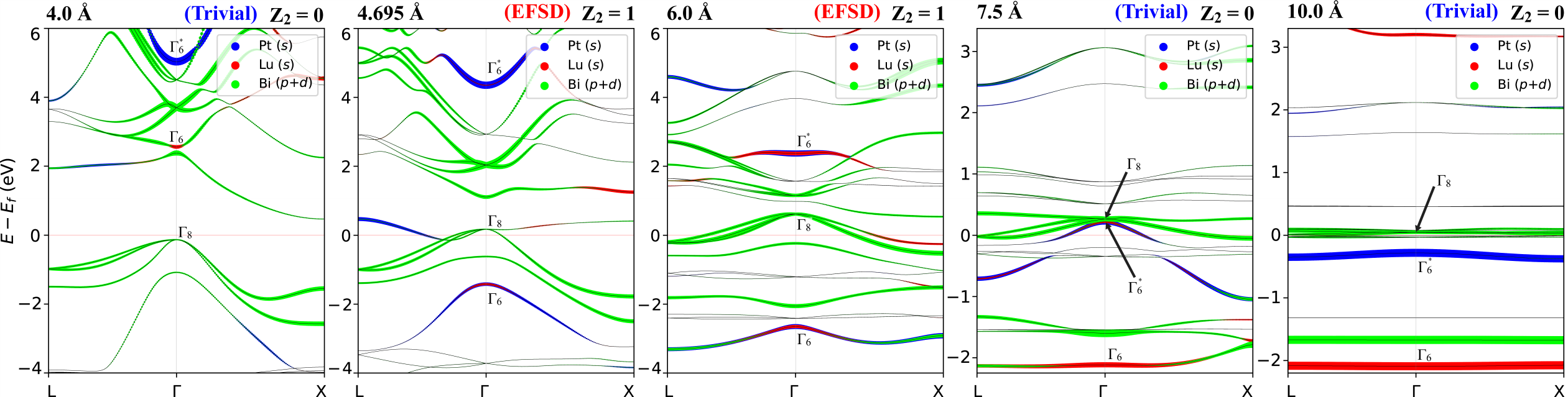}
    \caption{\label{fig:LuPtBi} Band structure with projection of the atomic orbital character, and topological classification for LuPtBi for different lattice parameters.}
\end{figure*}

\section{Results and discussion}
\label{sec:Results-discussion}

\begin{figure*}[t]
\includegraphics[width=1.8\columnwidth]{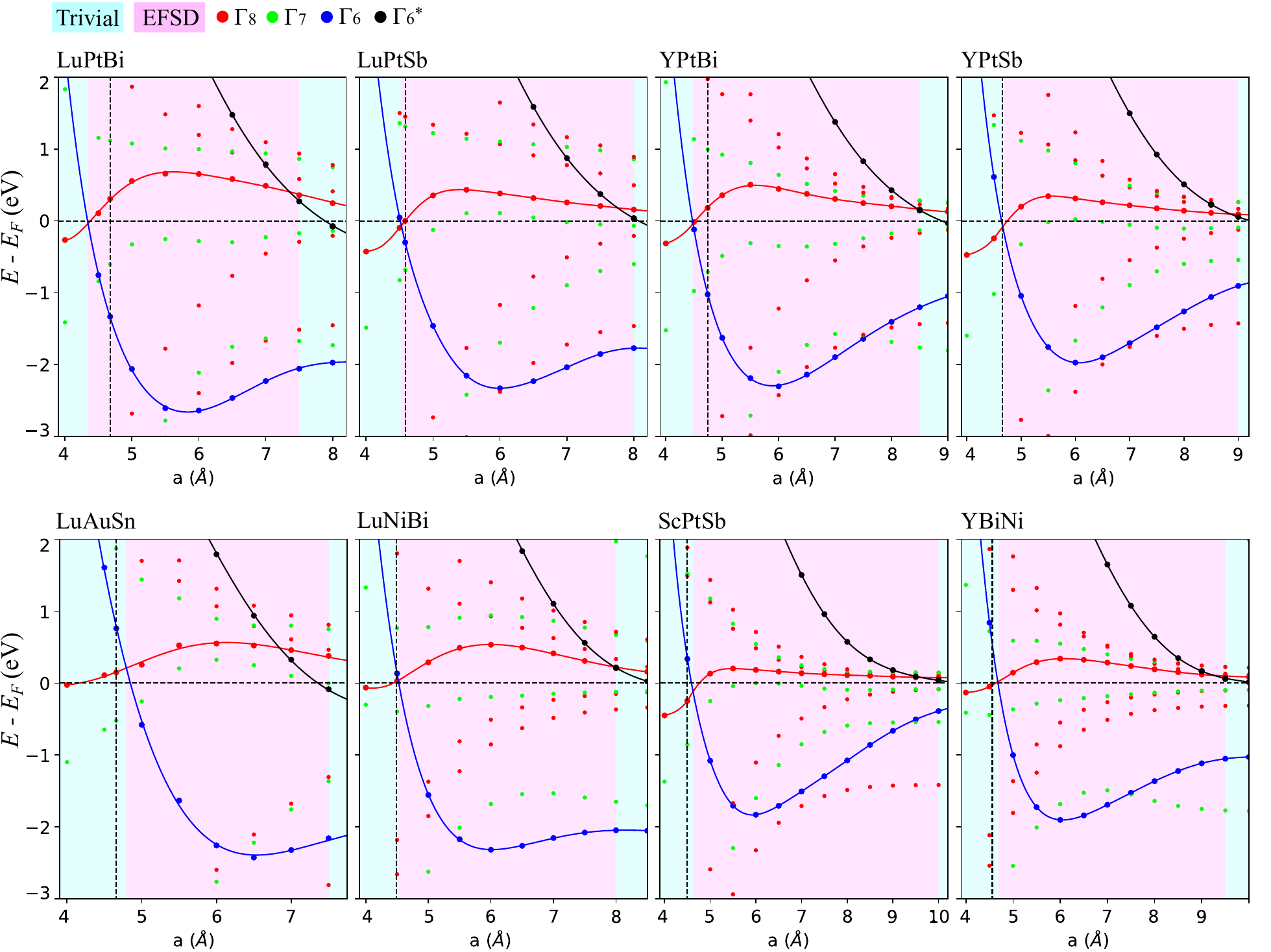}
    \caption{\label{fig:half-Heusler-irreps} Behavior of the IRREPs $\Gamma_6$, $\Gamma_8$ (over the $\Gamma$-point) as well as the topological classification (through the TQC approach), for a set of hH materials as a function of the increase/decrease of the lattice parameter.}
\end{figure*}

What motivated us to conduct this study was the unusual nature of half-Heusler materials in exhibiting transition to non-trivial topology when an increase in the lattice parameter is introduced into the system. This approach is opposite to the one commonly adopted to induce non-trivial topology in small-gap trivial materials. For example, let us cite the application of pressure on the bulk of IV-VI semiconductors PbTe/PbSe to obtain the 3D topological crystalline insulator phase \citep{IV-VI-transition-1,IV-VI-transition-2}. As reported in the literature \citep{chadov2010t,feng2010,lin2010,xiao2010} and observed by us in this study, half-Heusler compounds exhibit trivial topology when subjected to pressure and non-trivial topology under tension (see Figures \ref{fig:LuPtBi} and \ref{fig:half-Heusler-irreps}). A curious fact we found regarding the band ordering of half-Heusler compounds is that the system remains classified as non-trivial even for unrealistic expansion values. This led us to conduct the following heuristic analysis: If we continuously move the atoms in the lattice apart from each other through an adiabatic expansion process, no structural symmetry will be broken. However, the system must necessarily converge to the trivial phase when the interaction between the atoms becomes negligible. In this situation, we expect to observe the attainment of trivial character accompanied by the re-establishment of the trivial ordering between the $\Gamma_6$/$\Gamma_8$ states.

For this investigation, we calculated the electronic structure of a set of hH materials, varying the lattice parameter within the range of $4$ to $15$\,{\AA} in steps of $0.5$\,{\AA}, allowing us to map how the energy dispersion, atomic orbital character, irreducible representations (IRREPs), and topological classification of states evolve with increasing stress in the material. It should be noted that because the equilibrium lattice parameter of the analyzed compounds is around $4.5$\,{\AA} (vertical dashed lines in Figure \ref{fig:half-Heusler-irreps}), the range we used in this study also defines a small region of pressure application on the materials.

Since hH compounds have an electronic structure with semimetallic character, we decided to use the topological quantum chemistry (TQC) approach \citep{TQC1} for topological classification, which resulted in trivial phases and enforced semimetals with Fermi degeneracy (ESFD), indicated in Figures \ref{fig:LuPtBi} and \ref{fig:half-Heusler-irreps} for each material and configuration analyzed. In TQC, the classification of the ESFD phase as a topological state is based on a physical consequence, where the semimetallic state gives rise to the 3D topological insulator phase through structural symmetry breaking, capable of lifting the degeneracy at the Fermi level while preserving the inverted ordering of states (see Ref. \citep{TQC2} - Appendix J). This mechanism has been used in several works to verify the non-trivial topology of half-Heusler compounds \citep{lin2010, xiao2010, chadov2010t}. In addition to the TQC classification, we calculated the Z$_2$ topological index ($v_0$, $v_1$,$v_2$,$v_3$) \citep{kane2005z,roy2009topological} for the states near the Fermi level.

As expected, increasing the lattice parameter induces trivial systems to the non-trivial classification (EFSD). However, once the topological state is obtained, increasing the tension does not lead to the expected behavior (trivial phase accompanied by $\Gamma_6$/$\Gamma_8$ ordering). What we actually found is that as the tension increases, the $\Gamma_6$ state decreases in energy while $\Gamma_8$ remains above the Fermi level, such that $\Gamma_6$/$\Gamma_8$ ordering is always preserved. Interestingly, we found that the trivial phase is only truly re-established when a $\Gamma_6$ state originally located at high energies (which we will label as $\Gamma_6^*$) is shifted towards the Fermi level and crosses the $\Gamma_8$ state as lattice parameter increases. This result is reinforced by the confirmation of the trivial phase both by TQC and by the index Z$_2$ = 0 on the state $\Gamma_8$ (highest energy state and with non-zero occupation along the ZB).

The behavior described above occurs for all the analyzed hH materials, being better visualized in Figure \ref{fig:half-Heusler-irreps}, where we plot the energy eigenvalue of the states with the respective irreducible representation over the $\Gamma$-point as a function of the lattice parameter. This behavior demonstrates that the $\Gamma_6$/$\Gamma_8$ ordering relative to the Fermi level actually corresponds to a trivial ordering intrinsically linked to the energy levels of isolated atoms, while the $\Gamma_6^*$/$\Gamma_8$ ordering is what truly establishes the non-trivial character of this class of materials. Additionally, the $\Gamma_6$/$\Gamma_8$ ordering we expected to observe will only occur in the opposite direction, with a decrease in the lattice parameter, leading to the trivial phase due to a double band inversion process.

The origin of states can be understood by observing Figure \ref{fig:LuPtBi}, where we project the atomic orbital contributions, to the electronic structure of LuPtBi as a function of the lattice parameter variation. It can be seen that in the limit of isolated ions ($a \rightarrow \infty$), the states of interest show contribution predominant from the valence states of each ion, more precisely, $\Gamma_6$ ($s$-Lu ``site X''), $\Gamma_6^*$ ($s$-Pt ``site Y''), and $\Gamma_8$ ($p$-Bi ``site Z''). As the lattice parameter is reduced towards the equilibrium position, the $\Gamma_6$ state rises in energy, generating a repulsion process of the $\Gamma_6^*$ state to higher energies (behavior visualized in Figure \ref{fig:half-Heusler-irreps}), leading to the inverted ordering $\Gamma_6^*$/$\Gamma_8$. This effect can be better understood if we visualize the half-Heusler structure of LuPtBi as the introduction of site X (Lu) at the center of the zinc-blende YZ (PtBi) lattice, leading to crystal field splitting and a stronger interaction force among the elements of the lattice.

Notice here that in the atomic limit, the bands of character $\Gamma_6$ arise from the Lu-s and Pt-s orbitals that interact with each other, while the $\Gamma_8$ band arises from the Bi-p orbitals. We can understand the change in the band ordering from a simple model. Taking the most general $k \cdot p$ Hamiltonian for the orbitals composing the 2D $\Gamma_6$, $\Gamma_6^{*}$ and 4D $\Gamma_8$ irreducible representations, satisfying the hH symmetries is given by the Eq.~\ref{eq:kp-ham} Hamiltonian. At $k=0$ ($\Gamma$-point) such Hamiltonian is reduced to a simple three-level system with two interacting levels (each doubly degenerate) and a non-interacting one (4-fold degenerated), with k-independent interaction constants. That is the $V_{66}$ and $V_{66^*}$. Based on this analysis we can propose a phenomenological model connecting such constants with the system strain.

For instance, for the XYZ hH [X=Lu, Y, Sc; Y=Pt, Au, Ni; and Z=Bi, Sb, Sn], let the X-$s$ and Y-$s$ orbitals that present the same symmetry with energy $\epsilon_{X}$ and $\epsilon_{Y}$ interact with each other. Let $t$ be the interaction between first neighbor (1N) heteroatoms and $u$ the interaction between the same atoms from different unit cells (second neighbors, 2N). In order to capture the interaction in different strains, the couplings are chosen to decay exponentially with the distance (orbital-$s$ coupling) $t=t_0e^{-\alpha (x - 1/2)}$, $u=u_0 e^{-\beta (x - 1/\sqrt{2})}$, where $\alpha$ and $\beta$ are in units of the inverse of the lattice constant ($a_0^{-1}$), $x$ and the factors $1/2$ and $1/\sqrt{2}$ are the 1N and 2N distance. In the range of explored strain, the $\Gamma_8$ orbitals remained almost unchanged, and symmetry imposes it to be decoupled from the s orbitals, therefore we included it from the Z-$p$ orbital energy $\epsilon_{Z}$. Our effective Hamiltonian for $k$=0, considering the eight neighboring atoms reads
\begin{equation}
    H= \begin{pmatrix}
       \epsilon_{X} + 8u  & 8t & 0 \\
       8t & \epsilon_{Y} + 8u & 0 \\
       0 & 0 & \epsilon_{Z}
    \end{pmatrix}.
\end{equation}

\begin{figure}[t]
\includegraphics[width=1.0\columnwidth]{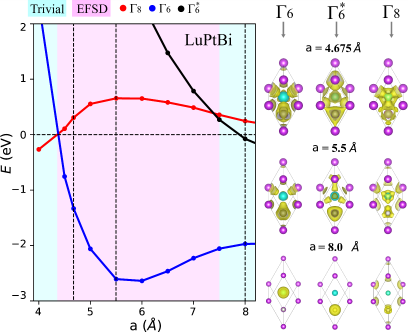}
    \caption{\label{fig:LuPtBi_Density} (left panel)Level evolution as a function of the lattice parameter. (right panel) Real space projection (LDOS) of the relevant levels, gray, blue and magenta are the Pt, Lu and Bi atoms respectively.}
\end{figure}

Here the eigenvalues are given by $E_1= \epsilon_Z$, $E_{2\pm} = \bar{\epsilon} + 8u \pm \sqrt{\bar{\delta}^2 + t^2}$, with $\bar{\epsilon} = (\epsilon_X + \epsilon_Y)/2$, and $\delta = (\epsilon_X - \epsilon_Y)/2$. Such a model captures the observed behavior present on the $\Gamma_8$/$\Gamma_6$/$\Gamma_6^{*}$ levels evolution. For instance, starting from the isolated system $t=u=0$ the states are related to the on-site energy of each orbital. Bringing the atoms together the first neighbor interactions start splitting the two $\Gamma_6$ states, from $8$\,{\AA} down to $6$\,{\AA} in the left panel of Fig.~\ref{fig:LuPtBi_Density}, and leading to the first crossing of the $\Gamma_6^*$ with the $\Gamma_8$ states. Further diminishing the lattice constant the second neighbor interaction ($u$) begins to take part in a shift of the on-site term, ruled by the diagonal terms in effective Hamiltonian, leading to the second crossing of the $\Gamma_6$ and $\Gamma_8$ states. Additionally, here we can extract an interpretation over the $\Gamma_6$ and $\Gamma_6^*$, being a bond and anti-bond-like states coming from the Pt/Lu-s orbitals, as captured by the model. In Figure \ref{fig:LuPtBi_Density} (right panels) we show the DFT calculation for the square modulus of each state wavefunction for different lattice parameters. Here, for further distant atoms ($a=8$\,{\AA}) the $\Gamma_6$/$\Gamma_6^*$/$\Gamma_8$ states are localized respectively on the Lu/Pt/Bi atoms. Upon bringing the atoms closer, the overlap wavefunction between the Lu/Pt atoms dictates a bonding and anti-bonding behavior.

\begin{figure}[h]
\includegraphics[width=1.0\columnwidth]{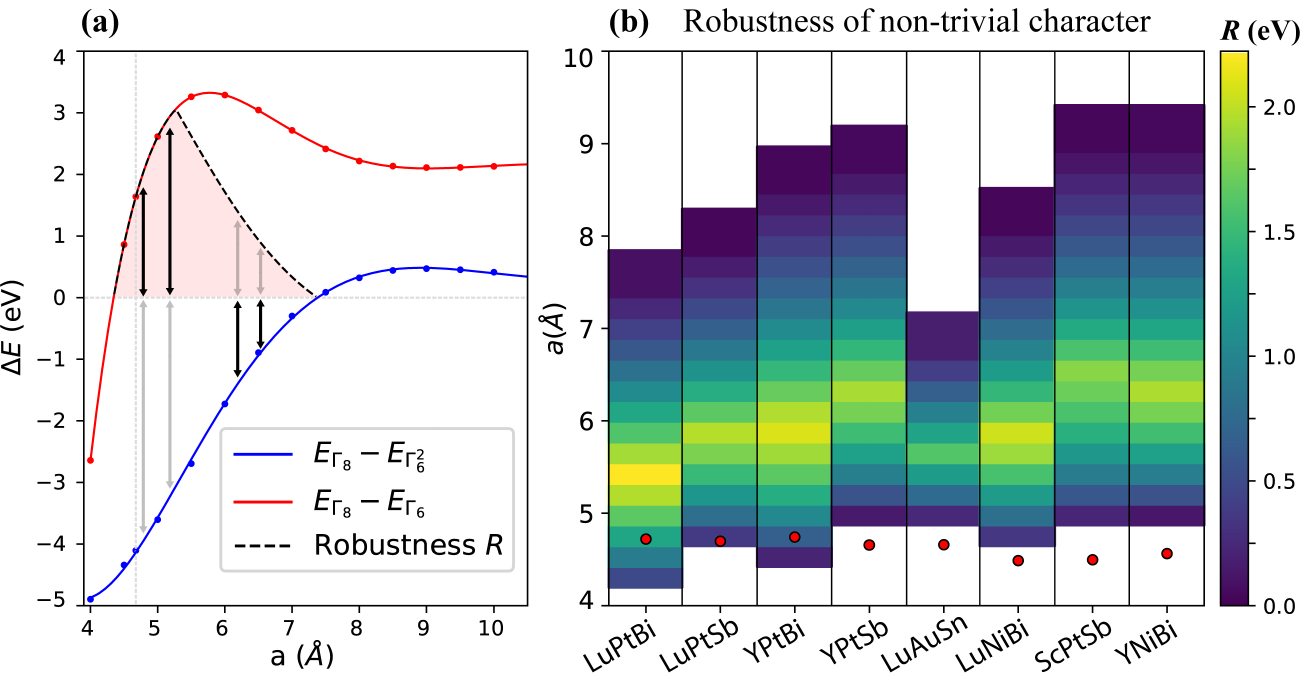}
    \caption{\label{fig:Robustness} (a) Energy gaps between the $\Gamma_6$/$\Gamma_6^*$and $\Gamma_8$ bands dictating the topologically non-trivial region when the Robustness (dashed lines) is greater than zero. (b) Robustness of the topologically non-trivial phase for the explored half-Heusler systems, the red dots mark the equilibrium lattice constant.}
\end{figure}

This repulsion of the $\Gamma_6$ and $\Gamma_6^*$ states not only rules the two topological transitions but also gives an energetic robustness for the topological band inversion. The system presents two relevant energy differences $\Delta_1 = E[\Gamma_8 ] - E[\Gamma_6 ]$ and $\Delta_2 = E[\Gamma_8 ] - E[\Gamma_6^* ]$, a negative value dictates that an inversion between the $\Gamma_8$ and the two $\Gamma_6$ in relation to the trivial configurations where $E[\Gamma_6 ]$ and $E[\Gamma_6^* ] < E[\Gamma_8 ]$. In Fig.~\ref{fig:Robustness}(a) we show the evolution of these energy differences for different lattice parameters. Overall the topological phase will be manifested if $\Delta_1 \Delta_2 < 0$. Perturbations and defects in topological systems that overcome the topological gap can lead to trivial phases \cite{PhysRevB.104.214206}. Here we can define the topological robustness as the minimal gap $R = \min \left\{ \Delta_1 ,\,\Delta_2 \right\}$ that a perturbation needs to overcome to induce a trivial phase, such minimum gap is shown in dashed lines Fig.~\ref{fig:Robustness}(a) for the LuBiPt. We have computed by DFT calculations the robustness for the explored half-Heusler systems, Fig.~\ref{fig:Robustness}(b). Here, the systems that are topologically non-trivial in the equilibrium geometry (LuBiPt, LuSbPt, and YBiPt) can have a more robust phase by a tensile strain. For instance, the minimum inverted gap ($R$) in LuBiPt can reach values of $2$\,eV for a $10\%$ strain, which has shown to be possible in half-Heusler membranes \cite{Du2023}. Additionally, the trivial systems (YSbPt, LuSbAu, LuBiNi, ScSbPt, and YBiNi) can have a trivial to non-trivial transition upon the same order of tensile strain.

\section{Conclusions}
\label{sec:conclusions}

In conclusion, through a heuristic procedure of expansion/contraction applied to half-Heusler compounds, we found that the topological classification of this class of materials is governed by a three-band mechanism between $\Gamma_6$, $\Gamma_8$, and $\Gamma_6^*$  states, providing a distinction in the mechanism governing zinc-blend III-V materials. The non-trivial phase of half-Heusler compounds is preserved based on the relative ordering of the $\Gamma_6$, $\Gamma_8$, and $\Gamma_6^*$ states at the Fermi level. Such mechanism provide uncommon robustness to the non-trivial phase of half-Heusler compounds under tension, reaching topological gaps up to $2$\,eV. We discuss the origin of such relevant states from the isolated atomic level and its manipulation with strain, giving a degree of design in the experimental construction of topological half-Heusler devices.

\begin{acknowledgments}

The authors acknowledge financial support from the Brazilian agencies FAPESP (grants 22/08478-6, 2023/12336-5, 19/20857-0, and 17/02317-2), CNPq (INCT - Materials Informatics and INCT - Nanocarbono), and LNCC - Laboratório Nacional de Computação Científica for computer time (projects ScafMat2 and EMT2D).

\end{acknowledgments}


\bibliography{bib}

\end{document}